\documentclass{ws-procs9x6}

\usepackage[normalem]{ulem} % \sout{old text} for strikeout
\usepackage[dvips]{color} % For blue in-text comments and additions
\renewcommand\sout{\bgroup \color{red} \ULdepth=-.5ex \ULset}

\begin{document}

\title{RECENT PROGRESS ON THE DETERMINATION OF THE SYMMETRY ENERGY}

\author{LIE-WEN CHEN$^{1,2}$}

\address{$^{1}$INPAC, Department of Physics and Shanghai Key Laboratory for Particle
Physics and Cosmology, Shanghai Jiao Tong University, Shanghai 200240, China\\
$^{2}$Center of Theoretical Nuclear Physics, National Laboratory of Heavy-Ion
Accelerator, Lanzhou, 730000, China\\
E-mail: lwchen@sjtu.edu.cn}

\begin{abstract}
We summarize the current status on constraining the density dependence of the
symmetry energy from terrestrial laboratory measurements and astrophysical
observations. While the value $E_{\text{\textrm{sym}}}({\rho _{0}})$ and density slope
$L$ of the symmetry energy at saturation density $\rho _{0}$ can vary largely
depending on the data or methods, all the existing constraints are essentially 
consistent with $E_{\text{\textrm{sym}}}({\rho _{0}}) = 31 \pm 2$ MeV and $L = 50 \pm 20$ MeV.
The determination of the supra-saturation density behavior of the symmetry energy
remains a big challenge.

\end{abstract}


\bodymatter

\section{Introduction}\label{aba:sec1}

In nuclear physics and astrophysics, there is currently of great interest to 
determine the density dependence of the nuclear symmetry energy $E_{\text{sym}}(\rho )$ 
that essentially characterizes the isospin dependent part of the equation of state 
(EOS) of asymmetric nuclear matter. The exact knowledge on the symmetry energy is 
important for understanding not only many problems in nuclear physics, such as the 
structure of radioactive nuclei, the reaction dynamics induced by rare isotopes, 
the liquid-gas phase transition in asymmetric nuclear matter, and the isospin evolution of QCD
phase diagram at finite baryon chemical potential, but also many critical issues
such as the properties of neutron stars and supernova explosion mechanism in
astrophysics~\cite{Lat04,Ste05,Bar05,CKLY07,LCK08}. The symmetry energy
may also be relevant to some interesting issues regarding possible new physics
beyond the standard model~\cite{Hor01b,Sil05,Wen09}. During the last decade,
although significant progress has been made both experimentally and
theoretically on constraining the density dependence of the symmetry
energy~\cite{Bar05,LCK08}, large uncertainties on $E_{\text{sym}}(\rho )$ still
exist, especially its super-normal density behavior remains elusive and largely
controversial~\cite{Xia09,Fen10,Rus11,XuC10b}. To reduce the uncertainties of 
the constraints on $E_{\text{sym}}(\rho )$ thus provides a strong motivation 
for studying isospin nuclear physics in radioactive nuclei at the new/planning 
rare isotope beam facilities around the world, such as CSR/Lanzhou and 
BRIF-II/Beijing in China, RIBF/RIKEN in Japan, SPIRAL2/GANIL in France, 
FAIR/GSI in Germany, FRIB/NSCL in USA, SPES/LNL in Italy, and KoRIA in Korea.

In the present talk, we summarize the current status on constraining the
density dependence of the symmetry energy from terrestrial laboratory
measurements and astrophysical observations, including nuclear reactions,
nuclear structures, and the properties of neutron stars.

\section{The symmetry energy}

\label{EOS}

The EOS of isospin asymmetric nuclear matter, given by its binding energy
per nucleon, can be expanded to $2$nd-order in isospin asymmetry $\delta $
as
\begin{equation}
E(\rho ,\delta )=E_{0}(\rho )+E_{\mathrm{sym}}(\rho )\delta ^{2}+O(\delta
^{4}),  \label{EOSANM}
\end{equation}%
where $\rho =\rho _{n}+\rho _{p}$ is the baryon density with $\rho _{n}$ and
$\rho _{p}$ denoting the neutron and proton densities, respectively; $\delta
=(\rho _{n}-\rho _{p})/(\rho _{p}+\rho _{n})$ is the isospin asymmetry; $%
E_{0}(\rho )=E(\rho ,\delta =0)$ is the binding energy per nucleon in
symmetric nuclear matter, and the symmetry energy is expressed as
\begin{equation}
E_{\mathrm{sym}}(\rho )=\frac{1}{2!}\frac{\partial ^{2}E(\rho ,\delta )}{%
\partial \delta ^{2}}|_{\delta =0}.  \label{Esym}
\end{equation}%
Neglecting the contribution from higher-order terms in Eq. (\ref{EOSANM}) 
leads to the well-known empirical parabolic law for the EOS of
asymmetric nuclear matter, which has been verified by all many-body theories
to date, at least for densities up to moderate values \cite{LCK08,Cai12}. As a
good approximation, the density-dependent symmetry energy $E_{\mathrm{sym}%
}(\rho )$ can thus be extracted from the parabolic approximation as
\begin{equation}
E_{\mathrm{sym}}(\rho )\approx E(\rho ,\delta =1)-E(\rho ,\delta =0). \label{EsymPA}
\end{equation}

Around the saturation density $\rho _{0}$, the nuclear symmetry energy $%
E_{\mathrm{sym}}(\rho )$\ can be expanded, e.g., up to 2nd-order
in density, as
\begin{eqnarray}
E_{\mathrm{sym}}(\rho ) &=&E_{\mathrm{sym}}(\rho _{0})+L\chi +\frac{K_{%
\mathrm{sym}}}{2!}\chi ^{2}+O(\chi ^{3}),  \label{EsymExpand}
\end{eqnarray}%
where $\chi =\frac{\rho -\rho _{0}}{3\rho _{0}}$ is a dimensionless variable
characterizing the deviations of the density from $\rho _{0}$, and 
$L$ and $K_{\mathrm{sym}}$ are the slope parameter and curvature parameter, respectively, i.e.,
\begin{eqnarray}
L =3\rho _{0}\frac{dE_{\mathrm{sym}}(\rho )}{\partial \rho }|_{\rho =\rho
_{0}},
K_{\mathrm{sym}} =9\rho _{0}^{2}\frac{d^{2}E_{\mathrm{sym}}(\rho )}{%
\partial \rho ^{2}}|_{\rho =\rho _{0}}.
\end{eqnarray}

\section{The symmetry energy around the saturation density}

During the last decade, a number of experimental probes have been proposed
to constrain the density dependence of the symmetry energy. Most of them are
for the symmetry energy around the saturation density while a few of probes
are for the supra-saturation density behaviors. In this section, we summarize
the present status on constraining the symmetry energy around the saturation
density, mainly, the parameters $E_{\text{\textrm{sym}}}(\rho _{0})$ and $L$,
from nuclear reactions, nuclear structures,
and the properties of neutron stars.

\subsection{Nuclear reactions}

Nuclear reactions, mainly including heavy ion collisions and nucleon-nucleus
scattering, provide an important tool to explore the density dependence of
the symmetry energy.

\subsubsection{Heavy ion collisions}

\begin{figure}[tbp]
\begin{center}
\includegraphics[scale=1.2]{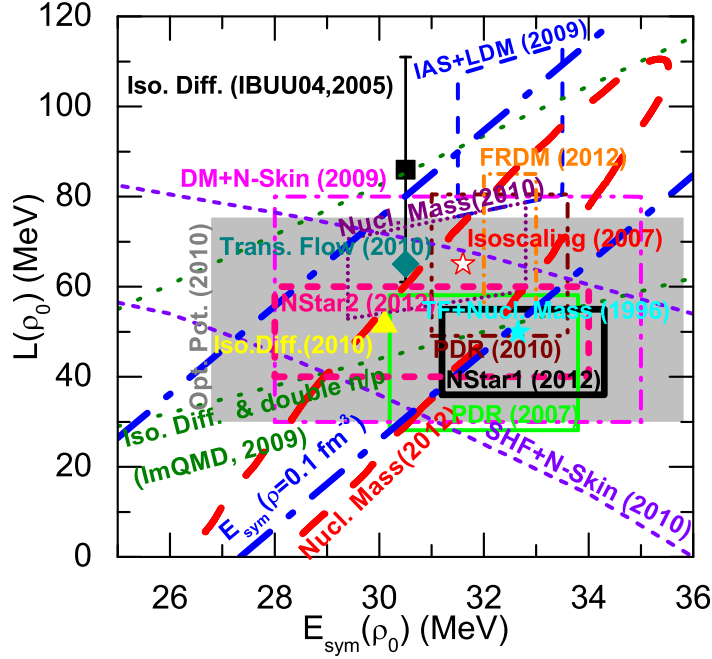}
\end{center}
\caption{(Color online) Constraints on $E_{\text{\textrm{sym}}}(\rho _{0})$ and $L$
from different experiments or methods. See text for details.}
\label{LEsym}
\end{figure}

One important progress on constraining the density dependence of the
symmetry energy is from the isospin dependent Boltzmann-Uehling-Uhlenbeck
(IBUU04) transport model analysis~\cite{Che05a} on the isospin diffusion
data from NSCL-MSU~\cite{Tsa04}. It is found that the degree of isospin diffusion in
heavy-ion collisions is affected by both the stiffness of the nuclear
symmetry energy and the momentum dependence of the nucleon potential.
Using a momentum dependence derived from
the Gogny effective interaction and the corresponding isospin dependent in-medium
nucleon-nucleon scattering cross sections, the experimental data from NSCL-MSU on
isospin diffusion leads to a constraint of $L = 86 \pm 25$ MeV with
$E_{\text{\textrm{sym}}}({\rho _{0}}) = 30.5$ MeV~\cite{Che05a,Che05b}, which is
shown as a solid square with error bar with a label ``\textbf{Iso. Diff. (IBUU04,2005)}"
in Fig.~\ref{LEsym}.  It should be mentioned that the constraint in the original
publication~\cite{Che05a,Che05b} is $L = 88 \pm 25$ MeV and
$E_{\text{\textrm{sym}}}({\rho _{0}}) = 31.6$ MeV, due to the application of
the parabolic approximation Eq.~(\ref{EsymPA}) for the symmetry energy. This
constraint is significantly softer than the prediction by transport model
simulation with momentum-independent interaction~\cite{Tsa04} and in agreement with
microscopic theoretical calculations.

The isoscaling of the fragment yields in heavy ion collisions has been shown
to be a good probe of the symmetry energy~\cite{Tsa01}. By analyzing the isoscaling of the
fragment yields in Ar+Fe/Ca+Ni, Fe+Fe/Ni+Ni, Ar+Ni/Ca+Ni, and
Fe+Ni/Ni+Ni reactions at Fermi energy region, within the antisymmetrized
molecular dynamic (AMD) model, a constraint of $E_{\text{\textrm{sym}}}({\rho _{0}}) = 31.6$
MeV and $L = 65 \pm 25$ MeV has been obtained in Ref.~\cite{She07}, which is denoted
by a open star with a label ``\textbf{Isoscaling (2007)}" in Fig.~\ref{LEsym}.

The double ratios of neutron and proton energy spectra in heavy ion collisions
provide a good probe of the symmetry energy. An improved quantum molecular
dynamics (ImQMD) transport model analysis~\cite{Tsa09} of the isospin diffusion data from
two different observables and the ratios of neutron and proton spectra in
collisions at E/A$=50$ MeV involving $^{112}$Sn and $^{124}$Sn nuclei has led to
a constraint on $E_{\text{\textrm{sym}}}({\rho _{0}})$ and $L$ at $95\%$ confidence
level, corresponding to $2$ standard deviations from the minimum $\chi^2$.
This constraint is denoted by the region between two dotted lines with a label
``\textbf{Iso. Diff. \& double n/p (ImQMD, 2009)}" in Fig.~\ref{LEsym}. A more recent
ImQMD model analysis~\cite{Sun10} of the isospin diffusion data from heavy ion collisions at lower
incident energy (E/A$=35$ MeV) involving $^{112}$Sn and $^{124}$Sn nuclei has led to a
constraint of $E_{\text{\textrm{sym}}}({\rho _{0}}) = 30.1$ MeV and $L = 51.5$ MeV,
which is shown by solid up-triangle with a label ``\textbf{Iso. Diff. (2010)}" in Fig.~\ref{LEsym}.

The isospin effects of fragment transverse flows in heavy ion collisions are
useful for extracting information on the symmetry energy. In a recent work~\cite{Koh10},
the transverse flow of intermediate mass fragments (IMFs) has been investigated for
the 35 MeV/u $^{70}$Zn +$^{70}$Zn, $^{64}$Zn + $^{64}$Zn, and $^{64}$Ni + $^{64}$Ni
systems. The analysis based on the AMD model with the GEMINI code treatment for
statistically de-excitation of the hot fragments leads to a constraint
$E_{\text{\textrm{sym}}}({\rho _{0}}) = 30.5$ MeV and $L = 65$ MeV, which is shown
by solid diamond with a label ``\textbf{Trans. Flow (2010)}" in Fig.~\ref{LEsym}.

\subsubsection{Nucleon optical potential}

Experimentally, there have accumulated a lot of data for elastic scattering
of proton (and neutron) from different targets at different beam energies
and (p,n) charge-exchange reactions between isobaric analog states. These data
provide the possibility to extract information on the isospin dependence of
the nucleon optical potential, especially the energy dependence of the nuclear
symmetry potential. Based on the Hugenholtz-Van Hove theorem, it has been 
shown recently~\cite{XuC10,XuC11,CheR12} that both 
$E_{\text{\textrm{sym}}}({\rho _{0}})$ and $L$ can be completely and analytically
determined by the nucleon optical potentials. Averaging all nuclear symmetry
potentials constrained by world data available in the literature since 1969
from nucleon-nucleus scatterings, (p,n) charge-exchange reactions, and
single-particle energy levels of bound states, the constraint
$E_{\text{\textrm{sym}}}({\rho _{0}}) = 31.3 \pm 4.5$ MeV and $L = 52.7 \pm 22.5$ MeV
are simultaneously obtained~\cite{XuC10}, and this constraint is indicated by
the gray band with a label ``\textbf{Opt. Pot. (2010)}" in Fig.~\ref{LEsym}.

\subsection{Nuclear structures}

In recent years, more and more constraints on the symmetry energy have been
obtained from the analyses of nuclear structure properties, such as the nuclear
mass (ground state binding energy), the neutron skin thickness, the nuclear
isobaric analog state energies, and pygmy dipole resonances. We summarize these
constraints in the following.

\subsubsection{Nuclear mass}

The nuclear mass data are probably the most accurate, richest, and least
ambiguous in the nuclear data library. The Thomas-Fermi model analysis~\cite{Mye96}
of 1654 ground state mass of nuclei with $N,Z \ge 8$ has given rise to
$E_{\text{\textrm{sym}}}({\rho _{0}}) = 32.65$ MeV and $L = 49.9$ MeV, which is
shown by solid star with a label ``\textbf{TF+Nucl. Mass (1996)}" in Fig.~\ref{LEsym}.

The symmetry energy coefficients $a_{\text{\textrm{sym}}}(A)$ of finite nuclei with
mass numbers $A = 20-250$ were determined from more than $2000$ precisely measured
nuclear masse~\cite{Liu10}. With the semiempirical connection between $a_{\text{\textrm{sym}}}(A)$
and the symmetry energy at reference densities, i.e.,
$E_{\text{\textrm{sym}}}({\rho _{A}})\approx a_{\text{\textrm{sym}}}(A)$, and
assuming a symmetry energy with density dependence of
$E_{\text{\textrm{sym}}}({\rho})=E_{\text{\textrm{sym}}}({\rho _{0}})(\rho /\rho_0)^{\gamma}$,
Liu \textit{et al}.~\cite{Liu10} obtained a constraint at $95\%$ confidence
level shown as a parallelogram with short-dotted-line sides in Fig.~\ref{LEsym},
labeled ``\textbf{Nucl. Mass(2010)}".

Within the Skyrme-Hartree-Fock (SHF) approach, it has been shown recently that a
value of  $E_{\text{\textrm{sym}}}({\rho _{A}})$ at a subsaturation reference
density $\rho _{A}$  leads to a positive linear correlation between
$E_{\text{\textrm{sym}}}({\rho _{0}} )$ and $L$~\cite{Che11}. Using recently extracted
$E_{\text{\textrm{sym}}}({\rho _{A}=0.1}$ {fm}$^{{-3}})
\approx a_{\text{\textrm{sym}}}(A=208) = 20.22-24.74$ MeV at $95\%$ confidence
level from more than 2000 measured nuclear masses, Chen~\cite{Che11} obtained
a constraint denoted by the region between two thick dash-dotted lines with a
label ``\textbf{$E_{\text{\textrm{sym}}}({\rho _{A}=0.1}$ {fm}$^{{-3}})$(2011)}"
in Fig.~\ref{LEsym}.

The finite-range droplet(FRDM) model has been shown to be very successful to
describe the nuclear ground state mass. The parameters in the macroscopic droplet
part of the FRDM model are related to the properties of the equation of state.
Using the new, more accurate FRDM-2011a version, Moller \textit{et al}.~\cite{Mol12}
analyzed the nuclear mass of the 2003 Atomic Mass Evaluation (AME2003), and
obtained the constraint $E_{\text{\textrm{sym}}}({\rho _{0}}) = 32.5 \pm 0.5$
MeV and $L = 70 \pm 15$ MeV shown as a square box bounded by short-dash-dotted
lines in Fig.~\ref{LEsym}, labeled ``\textbf{FRDM (2012)}".

In a more recent work~\cite{Lat12}, Lattimer and Lim used the confidence ellipse
method for nuclear mass fitting. Based on a SHF energy-density functional for
nuclear masses, they obtained a $95\%$ confidence ellipse for the
$E_{\text{\textrm{sym}}}({\rho _{0}})$-$L$ constraints shown by the thick dashed
lines in Fig.~\ref{LEsym}, labeled ``\textbf{Nucl. Mass(2012)}".

\subsubsection{Neutron skin thickness}

Theoretically, it has been established~\cite{Bro00,Che05b} that the neutron skin
thickness of heavy nuclei, given by the difference of their neutron and proton
root-mean-squared radii, provides a good probe of $E_{\text{\textrm{sym}}}(\rho )$.
The droplet model analyses~\cite{Cen09} on the neutron skin sizes measured in 26
antiprotonic atoms along the mass table leads to the constraint
$E_{\text{\textrm{sym}}}({\rho _{0}}) = 28-35$ MeV and $L = 30-80$ MeV shown as a
square box bounded by dash-dotted lines in Fig.~\ref{LEsym}, labeled ``\textbf{DM+N-Skin (2009)}".

A microscopic SHF analysis~\cite{Che10} on the neutron skin thickness of Sn isotopes has led to
a set of constraints corresponding to $95\%$ confidence levels, shown as a region
bounded by two short-dashed curves in Fig.~\ref{LEsym}, labeled ``\textbf{SHF+N-Skin (2010)}".

\subsubsection{Nuclear isobaric analog state energies}

The nuclear isobaric analog state (IAS) energies are believed to provide a
particularly clean and useful probe of the symmetry energy since the ambiguities
in the determination of the symmetry energy of finite nuclei from binding
energies caused by the Coulomb term can be removed~\cite{Dan03}. By fitting the
available data on the IAS and using the droplet surface symmetry energy,
Danielewicz and Lee~\cite{Dan09} obtained the constraint shown as a parallelogram
bounded by dashed lines in Fig.~\ref{LEsym}, labeled ``\textbf{IAS+LDM (2009)}".

\subsubsection{Pygmy dipole resonance}

The experimentally observed pygmy dipole (E1) strength might play an
equivalent role as the neutron rms radius in constraining the
symmetry energy~\cite{Pie06}. Excess neutrons forming the skin give rise to
pygmy dipole transitions at excitation energies below the giant
dipole resonance, and such transitions could represent a collective
vibration of excess neutrons against an isospin symmetric core.
Comparing the measured pygmy dipole strength in $^{130,132}$Sn to
that obtained within a relativistic mean-field approach, Klimkiewicz \textit{et al}.~\cite{Kli07}
obtained the constraint $E_{\text{\textrm{sym}}}({\rho _{0}}) = 30.2-33.8$ MeV
and $L = 28.1-58.1$ MeV shown as a square box bounded by think solid lines
in Fig.~\ref{LEsym}, labeled ``\textbf{PDR (2007)}". Another analysis~\cite{Car10} on
the measured pygmy dipole strength in $^{68}$Ni and $^{132}$Sn within the
relativistic and  non-relativistic mean-field approaches leads to the
constraint $E_{\text{\textrm{sym}}}({\rho _{0}}) = 31.0-33.6$ MeV
and $L = 49.1-80.5$ MeV shown as a square box bounded by dash-dot-dotted
lines in Fig.~\ref{LEsym}, labeled ``\textbf{PDR (2010)}".

\subsection{The properties of neutron stars}

Astrophysical observations of neutron star masses and radii provide important
probe for the equation of state of neutron-rich matter. In particular, neutron
star radii are strongly correlated with neutron matter pressures around the
saturation density~\cite{Lat01}.

In a recent work~\cite{Ste12}, Steiner and Gandolfi demonstrated that currently
available neutron star mass and radius measurements provide a significant
constraint on the EOS of neutron matter. Using a phenomenological parametrization
for EOS of neutron matter near and above the saturation density with partial
parameters determined by the quantum Monte Carlo calculations, they obtained a
constraint of $E_{\text{\textrm{sym}}}({\rho _{0}}) = 31.2-34.3$ MeV
and $L = 36-55$ MeV at $95\%$ confidence level based on Bayesian
analysis~\cite{Ste12,Tsa12}, and this constraint is shown as a square box bounded
by thick solid lines in Fig.~\ref{LEsym}, labeled ``\textbf{NStar1 (2012)}".
More recently, Lattimer and Lim~\cite{Lat12} performed a similar Bayesian analysis
of the available neutron star mass and radius measurements, they obtained a
constraint of $E_{\text{\textrm{sym}}}({\rho _{0}}) = 28-34$ MeV and
$L = 40-60$ MeV shown as a square box bounded by thick short-dahsed lines in
Fig.~\ref{LEsym}, labeled ``\textbf{NStar2 (2012)}".

Besides neutron star mass and radius, other properties of neutron stars may
also put constraints on the symmetry energy. For example, the binding energy of
neutron stars~\cite{New09}, the frequencies of torsional crustal
vibrations~\cite{Ste09,Gea11} and the r-model instability window~\cite{Wen12}
all consistently favor $L$ values less than about $70$ MeV.

\subsection{Discussions}

In Fig.~\ref{LEsym}, we include totally $18$ constraints on $L$ and
$E_{\text{\textrm{sym}}}({\rho _{0}})$ described above. Obviously, it cannot be
that all the constraints are equivalently reliable since some constraints do not
have overlap. It should be stressed that the symmetry energy cannot be measured
directly and each constraint shown in Fig.~\ref{LEsym} is based on a certain
theoretical model with some approximations or special assumptions.

\begin{figure}[tbp]
\begin{center}
\includegraphics[scale=0.9]{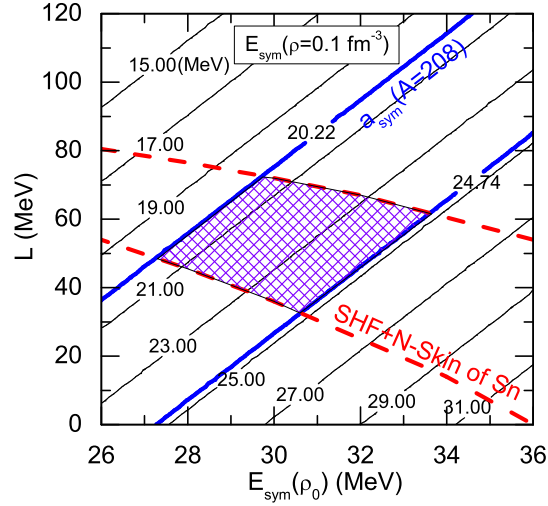}
\end{center}
\caption{(Color online) Contour curves in the $E_{\text{\textrm{sym}}}(\protect\rho _{0})$
-$L$ plane for $E_{\text{\textrm{sym}}}(\protect\rho =0.1$ fm$^{-3})$ from SHF calculations.
The region between the two thick solid lines represents the constraint obtained
with 20.22 MeV $\leq E_{\text{\textrm{sym}}}(\protect\rho _{A}=0.10\ $ fm$^{-3})$
$\leq $24.74 MeV while the region between the two thick dashed lines is the constraint
from the SHF analysis of neutron skin data of Sn isotopes within a $2\protect\sigma $
uncertainty \protect\cite{Che10}. The shaded region represents the overlap of the two
constraints. Taken from Ref.~\cite{Che11}.}
\label{Esym01EsymL}
\end{figure}

We would like to highlight two constraints in Fig.~\ref{LEsym}, i.e.,
``\textbf{SHF+N-Skin (2010)}" and
``\textbf{$E_{\text{\textrm{sym}}}({\rho _{A}=0.1}$ {fm}$^{{-3}})$(2011)}" since
both constraints are based on the same SHF analysis with $95\%$ confidence.
The two constraints are re-plotted in Fig.~\ref{Esym01EsymL}.
It is very interesting to see that while the constraint
``\textbf{$E_{\text{\textrm{sym}}}({\rho _{A}=0.1}$ {fm}$^{{-3}})$(2011)}"
indicates a linear positive correlation between $L$ and $E_{\text{\textrm{sym}}}({\rho _{0}})$,
the constraint ``\textbf{SHF+N-Skin (2010)}" displays a negative correlation.
Actually, only the constraint ``\textbf{SHF+N-Skin (2010)}"
among the $18$ constraints displays such negative correlation. This interesting feature makes
the constraint ``\textbf{SHF+N-Skin (2010)}" particularly important as combing it
with other constraints will significantly improve the constraint on $L$ and
$E_{\text{\textrm{sym}}}({\rho _{0}})$. It is interesting to see that the overlap of
``\textbf{SHF+N-Skin (2010)}" and
``\textbf{$E_{\text{\textrm{sym}}}({\rho _{A}=0.1}$ {fm}$^{{-3}})$(2011)}" is consistent
with all the other constraints shown in Fig.~\ref{LEsym} except ``\textbf{IAS+LDM (2009)}".
The latter neglected the higher-order density curvature contribution of the
symmetry energy and its inclusion may reduce the $L$ value~\cite{Lat12} (See also Ref.~\cite{Che11}).

The positive correlation between $L$ and $E_{\text{\textrm{sym}}}({\rho _{0}})$ from
``\textbf{$E_{\text{\textrm{sym}}}({\rho _{A}=0.1}$ {fm}$^{{-3}})$(2011)}" has been
clearly demonstrated in Ref.~\cite{Che11}, This feature implies that nuclear mass fitting
should lead to positive correlation between $L$ and $E_{\text{\textrm{sym}}}({\rho _{0}})$
since $E_{\text{\textrm{sym}}}({\rho _{A}=0.1}$ {fm}$^{{-3}})$ reflects the
symmetry energy of finite nuclei, which is demonstrated by the nice agreement between
the constraints ``\textbf{$E_{\text{\textrm{sym}}}({\rho _{A}=0.1}$ {fm}$^{{-3}})$(2011)}"
and ``\textbf{Nucl. Mass(2012)}".

The negative correlation between $L$ and $E_{\text{\textrm{sym}}}({\rho _{0}})$ from
``\textbf{SHF+N-Skin (2010)}" can be understood from the fact that the neutron skin
thickness is determined by the neutron and proton pressure difference at sub-saturation
density, namely, the density slope of symmetry energy at sub-saturation density
(rather than $L$), which increases with both $L$ and
$E_{\text{\textrm{sym}}}({\rho _{0}})$~\cite{Che12}.

From Fig.~\ref{LEsym} and Fig.~\ref{Esym01EsymL}, we can see that while
$E_{\text{\textrm{sym}}}({\rho _{0}})$ and $L$ can vary largely
depending on the data or methods, all the constraints are essentially
consistent with $E_{\text{\textrm{sym}}}({\rho _{0}}) = 31 \pm 2$ MeV and
$L = 50 \pm 20$ MeV.

\section{The symmetry energy at supra-saturation densities}

While significant progress has been made on constraining the symmetry energy
around the saturation density, the supra-saturation density behavior of the
symmetry energy remains elusive and largely controversial. FOPI data on the
$\pi^- /\pi^+$ ratio in central heavy-ion collisions at SIS/GSI energies favor
a quite soft symmetry energy at $\rho \ge 2\rho_0$ from the isospin and momentum
dependent IBUU04 model analysis~\cite{Xia09} while an opposite conclusion has
been obtained from the improved isospin dependent quantum molecular dynamics
(ImIQMD) model analysis~\cite{Fen10}. It should be mentioned that the ImIQMD
model analysis did not consider the energy dependent symmetry potential and
it cannot explain qualitatively the isospin fractionation phenomenon observed
in heavy ion collisions~\cite{Xu00,LiBA02}. A further careful check is
definitely needed to understand the model dependence.

In a more recent work, Russotto \textit{et al}.~\cite{Rus11} analyzed the elliptic-flow
ratio of neutrons with respect to protons or light complex particles from the
existing FOPI/LAND data for $^{197}$Au + $^{197}$Au collisions at 400 MeV/nucleon
within the UrQMD model, and they obtained a moderately soft symmetry energy with a
density dependence of the potential term proportional to
$(\rho/\rho_0)^{\gamma}$ with $\gamma = 0.9 \pm 0.4$.

Besides using heavy ion collisions to constrain the supra-saturation density
behavior of the symmetry energy, it has been proposed recently~\cite{Che11b} that the three
bulk characteristic parameters $E_{\text{\textrm{sym}}}({\rho _{0}})$, $L$ and
$K_{\text{\textrm{sym}}}$ essentially determine the symmetry energy with the density
up to about $2\rho_0$. This opens a new window to constrain the supra-saturation
density behavior of the symmetry energy from its density behaviors at the saturation
density.

\section{Summary}

Significant progress has been made both experimentally and theoretically on
constraining the density dependence of the symmetry energy after more than
one decade of studies in the community. Although the values of
$E_{\text{\textrm{sym}}}({\rho _{0}})$ and $L$ can vary largely
depending on the data or methods, all the constraints obtained so far from
nuclear reactions, nuclear structures, and the properties of neutron stars
are essentially consistent with $E_{\text{\textrm{sym}}}({\rho _{0}}) = 31 \pm 2$ MeV
and $L = 50 \pm 20$ MeV. More high quality data and more accurate theoretical
methods are needed to further reduce the theoretical and experimental
uncertainties of the constraints on $E_{\text{sym}}(\rho )$ around the saturation
density. In contrast, the determination of the supra-saturation density
behavior of the symmetry energy is still largely controversial and
remains a big challenge in the community.

\section*{Acknowledgments}
The author would like to thank Bao-Jun Cai, Rong Chen, Peng-Cheng Chu,
Wei-Zhou Jiang, Che Ming Ko, Bao-An Li, Kai-Jia Sun, Rui Wang, Xin Wang,
De-Hua Wen, Zhi-Gang Xiao, Chang Xu, Jun Xu, Gao-Chan Yong, Zhen Zhang,
and Hao Zheng for fruitful collaboration and stimulating discussions.
This work was supported in part by the NNSF of China under Grant
Nos. 10975097 and 11135011, the Shanghai Rising-Star Program under
grant No. 11QH1401100, the ``Shu Guang" project supported by
Shanghai Municipal Education Commission and Shanghai Education
Development Foundation, the Program for Professor of Special
Appointment (Eastern Scholar) at Shanghai Institutions of Higher
Learning, the Science and Technology Commission of Shanghai
Municipality (11DZ2260700), and the National Basic Research Program
of China (973 Program)under Contract No. 2007CB815004.

\end{document}